\documentstyle[12pt]{article}
\newcommand{\nn}{\nonumber\\}\newcommand{\p}[1]{(\ref{#1})}

\def\a{\alpha}\def\b{\beta}\def\g{\gamma}\def\d{\delta}

\def\k{\kappa}\def\l{\lambda}\def\S{\Sigma}
\def\th{\theta}\def\om{\omega}\def\Om{\Omega}
\begin{document}
\renewcommand{\thefootnote}{\fnsymbol{footnote}}
\thispagestyle{empty}
\begin{flushright}
HUB-EP-99/32, \\ 
TUW/99-15, \\ 
MPI-PhT99/28\\
hep-th/99mmnnn
\end{flushright}

\begin{center}
{\Large \bf $OSp$ supergroup manifolds, superparticles and supertwistors}

\bigskip
Igor Bandos \footnote{Lise Meitner Fellow.
On leave of absence from  Institute for Theoretical Physics
NSC Kharkov Institute of Physics and Technology
310108 Kharkov, Ukraine}

\bigskip
{\it  Institute for Theoretical Physics, 
Technical University of Vienna\\
Wiedner Haupstrasse 8-10, 
A-1040 Wien, Austria\\}
e-mail: bandos@tph32.tuwien.ac.at

\bigskip
Jerzy Lukierski\footnote{Supported in part by { KBN} grant
{ 2P03B13012}} 

\bigskip
{\it Institute for Theoretical Physics 
University of Wroclaw, \\
50-204 Wroclaw, Poland\\}
e-mail: lukier@proton.ift.uni.wroc.pl\\
and\\
{\it 
 Max Plank Institut f\"ur Physik (Werner-Heisenberg-Institute)\\
Fohringer Ring 6, D-80805 M\"unchen, Germany}

\bigskip
Christian Preitschopf and Dmitri Sorokin\footnote{A. von Humboldt Fellow.
On leave of absence from  Institute for Theoretical Physics,
NSC Kharkov Institute of Physics and Technology,
310108 Kharkov, Ukraine}

\bigskip
{\it Humboldt-Universit\"at zu Berlin,
Institut f\"ur Physik, \\
Invalidenstrasse 110, D-10115 Berlin, Germany}\\
e-mail: preitsch,sorokin@physik.hu-berlin.de

\bigskip

\setcounter{page}0
\newpage 

{\bf Abstract}
\end{center}
We construct simple twistor--like actions describing superparticles 
propagating on a coset superspace $OSp(1|4)/SO(1,3)$ (containing
the $D=4$ anti--de--Sitter space as a bosonic subspace),
on a supergroup manifold $OSp(1|4)$ and, generically, on $OSp(1|2n)$. 
Making two different contractions of the superparticle model
on the $OSp(1|4)$ supermanifold we get massless
superparticles in Minkowski superspace without and with tensorial 
central charges. 

Using a suitable parametrization of $OSp(1|2n)$ we obtain even 
$Sp(2n)$--valued Cartan forms which are quadratic in Grassmann 
coordinates of $OSp(1|2n)$.
This result may simplify the structure of brane actions in super--AdS 
backgrounds. For instance, the twistor--like actions constructed with 
the use of the even $OSp(1|2n)$ Cartan forms as supervielbeins are 
quadratic in fermionic variables.

We also show that the free bosonic twistor particle action describes 
massless particles propagating in arbitrary space-times with a conformally 
flat metric, in particular, in Minkowski space and AdS space. 
Applications of these results to the theory of higher spin fields and
to superbranes in AdS superbackgrounds are mentioned.

\renewcommand{\thefootnote}{\arabic{footnote}}
\setcounter{page}0
\newpage
\section{Introduction}
Conformal (super)symmetry plays an important role in
the theory of fundamental interactions based on  field--theoretical models
as well as on the theory of fundamental extended objects (strings etc.). 
Conformal geometrical structure allows one to replace
space--time geometry by twistor geometry, where
twistors are fundamental conformal spinors ($SU(2,2)$ spinors for
$D =4$) and space--time variables 
become twistor composites \cite{twistor}. Such a construction allows for
a supersymmetric extension \cite{fer} where
superspace variables are replaced by primary supertwistor coordinates
($SU(2,2|N)$ supertwistors in $D=4$ case).
 
In this paper we shall discuss twistors 
describing anti--de--Sitter (AdS) geometry.

The isometry of the AdS (super)spaces acts
on the (super)AdS boundary as the group of (super)conformal
transformations, and, therefore,
provides the group--theoretical basis for the AdS/CFT (conformal field 
theory) correspondence conjecture \cite{ads}
which attracted great deal of attention
during the last two years (see \cite{adsrev} for an exhaustive list of 
references).
On the other hand the AdS space is the one where higher spin fields may
nontrivially interact with each other \cite{fv}. 
In some aspects the technique
developed for the description of the theory of higher spin fields in
Minkowski \cite{misha1} and AdS spaces \cite{misha2}
resembles the (super)twistor approach \cite{misha3}.

In this respect it is tempting to look for possible links between these
different manifestations of conformal symmetry, AdS spaces, twistors and
their supersymmetric generalizations.

Motivated by the problem of finding a simple form of the action
for a superstring propagating in the $AdS_5\times S^5$ superbackground 
\cite{claus1},
in recent papers \cite{claus2} a  massive bosonic
twistor particle model in an $AdS_5$ space has been proposed  
and its classical and quantum conformal symmetry properties 
have been considered.

In \cite{1/4,bls} an $OSp(1|8)$ supertwistor model has been proposed for the 
description of a $D=4$ massless superparticle with the 
infinite spectrum of quantum states being described by
fields of arbitrary integer and half integer spins. The spin degrees of
freedom of the superparticle have been found to be associated with
six tensorial central charges which can extend the 
$N=1$, $D=4$ super--Poincare algebra.
It is natural to assume that the superparticle with tensorial central 
charges and the particle on the AdS space are different truncations of
the dynamics of a superparticle propagating in the supergroup manifold
of the isometries of the corresponding AdS superspace \cite{bls}.

An aim of this paper is to construct such a model in the supergroup 
manifold $OSp(1|4)$
and demonstrate how it is related to the $D=4$ twistor superparticle model
of higher spins \cite{1/4,bls}, and to a superparticle on the $AdS_4$ 
superspace $OSp(1|4)/SO(1,3)$.

Another motivation of this study is to find a way of constructing
simple worldvolume actions describing the dynamics of superbranes propagating
in AdS superbackgrounds, i.e. to make a progress in solving a vital AdS/CFT
correspondence problem \cite{2bs}--\cite{zhou}.

Using a suitable parametrization of $OSp(1|2n)$ 
(where $n$ is a natural number) we have found a simple
form of the even $OSp(1|2n)$ Cartan forms. They are only quadratic 
in Grassmann coordinates. This has allowed us to construct
simple actions quadratic in fermions for superparticles propagating 
on $OSp(1|4)/SO(1,3)$, $OSp(1|4)$ and, generically, on $OSp(1|2n)$.

The most interesting examples of the $OSp(1|2n)$ supergroups seem  
to be  $OSp(1|32)$  and  $OSp(1|64)$. 
In \cite{van,ferr,bars} \footnote{ 
An ${OSp(1|64)}$--invariant superparticle--like 
model has been discussed in \cite{bars} 
(see also \cite{bars2} and references therein).}
it has been shown that $OSp(1|32)$ and $OSp(1|64)$ contain the supergroup
structures of $D=11$ M--theory and $D=10$ superstrings.
In particular, $OSp(1|32)$ and $OSp(1|64)$ are extensions of the supergroups
$SU(2,2|4)$, $OSp(8|4)$ and $OSp(2,6|4)$ 
\footnote{We should remark that the notation $OSp(6,2|4)$ 
is a somewhat confusing name for the 
$AdS_7$ quaternionic supergroup described, 
in a complex parametrization, 
by the intersection of two complex supergroups $SU(4,4|4)$ and $OSp(8|4;C)$, 
with bosonic sectors being, respectively, the spinorial covering of $O(6,2)$
(space--time) and the spinorial covering $Sp(2;H)=USp(4;C)$ of $O(5)$ 
(the internal sector).} which are isometries of, respectively, 
$AdS_5\times S^5$, $AdS_4\times S^7$ and $AdS_7\times S^5$ superspaces. 
Reducing $OSp(1|32)$ and $OSp(1|64)$ down to the AdS--supergroups
one may hope to get simpler expressions for the Cartan forms of the latter, 
which might simplify the structure of actions for branes in corresponding
AdS superbackgrounds \cite{2bs}--\cite{zhou} \footnote{The Cartan forms
on supercosets of $SU(2,2|N)$ relevant to the construction of brane
actions on AdS superbackgrounds were calculated in early 80-ths \cite{abz}.}.

The plan of the paper is the following.

In Section 2 we review properties of the twistor formulation of bosonic
particle mechanics and demonstrate that the  single
twistor particle action  generically describes
particles propagating in arbitrary space--times which admit a conformally
flat metric, such as flat Minkowski space and the AdS space.

In Subsection 3.1 we consider the supertwistor description of a massless 
superparticle in flat $N=1$, $D=4$ superspace, 
and in Subsection 3.2 we construct
a twistor--like action for the description of the dynamics of a superparticle
in the super--AdS space $OSp(1|4)/SO(1,3)$. The action has a simple 
quadratic form in fermions and, hence, it should not be hard to perform its
quantization. However, in contrast to the bosonic
$AdS_4$ superparticle we have not managed to find a complete supertwistor
version of this model. 

In Section 4 we construct a twistor--like action for a superparticle 
on the supergroup manifold $OSp(1|4)$.  
This action is also quadratic in fermions, and 
upon an appropriate truncation it reduces to the models of 
Section 3. 

In Section 5 we describe a superparticle propagating on $OSp(1|2n)$ and
show that it preserves $2n-1$ supersymmetries associated with
Grassmann generators of $OSp(1|2n)$.

The $OSp(1|4)$ superalgebra and its Cartan forms required for the construction
of the actions are given in Appendix 1. In particular, we present
a simple form of the super--$AdS_4$ supervielbeins and spin connection
which are polynomials of only the second down in Grassmann coordinates.

In Appendix 2 we present Cartan forms of the supergroup $OSp(1|2n)$ which
can be made quadratic in Grassmann variables by an appropriate rescaling
of the latter.

\section{Twistor--like bosonic particles}
We start by recalling the form of an action for massless $D=4$ particles 
which serves as a dynamical basis for the transform from the 
space--time to the twistor description. The action is
\begin{equation}\label{1}
S=\int d\tau \lambda^A(\sigma_{m})_{A\dot A}
\bar\lambda^{\dot A}{d\over{d\tau}}x^m(\tau)=
\int \lambda\sigma_{m}\bar\lambda dx^m(\tau),
\end{equation}
where $x^m(\tau)$ $(m=0,1,2,3)$ is a particle trajectory in $D=4$ Minkowski
space, $\lambda^A(\tau)$ is a commuting two--component Weyl spinor and 
$\sigma^m_{A\dot A}=\bar\sigma^m_{\dot A A}$ are the Pauli matrices.

 From \p{1} we derive that the canonical conjugate momentum of $x^m(\tau)$ is
\begin{equation}\label{2}
p_m=\lambda\sigma_{m}\bar\lambda \quad \Rightarrow \quad
p_{A\dot A}={1\over 2}p_m\sigma^m_{A\dot A}
=\lambda_A\bar\lambda_{\dot A},
\end{equation}
whose square is identically zero since 
$\lambda^A\lambda^A\epsilon_{AB}\equiv 0$ 
\footnote{The two--component spinor indices are raised and lowered by the unit
antisymmetric matrices $\epsilon^{AB}=\epsilon_{AB}$, 
$\epsilon^{\dot A\dot B}=\epsilon_{\dot A\dot B}$.}, i.e.
\begin{equation}\label{03}
p_mp^m=0.
\end{equation}
We therefore conclude that the particle is massless.

In the action \p{1} we can make the change of variables by introducing the 
second commuting spinor
\begin{equation}\label{3}
\bar\mu_{\dot A}=i\lambda^A(\sigma_m)_{A\dot B} x^m\equiv
i\lambda^A x_{A\dot A}
\end{equation}
and its complex conjugate
\begin{equation}\label{3a}
\mu_{A}= -ix_{A\dot B}\bar\lambda^{\dot B}.
\end{equation}

The four--component spinors
\begin{equation}\label{4}
Z_{\alpha}=(\lambda_A, \bar\mu^{\dot A}), \quad
\bar Z^{\alpha}=(\mu^A, \bar\lambda_{\dot A})
\end{equation}
are called twistors.

In terms of \p{4} the action \p{1} takes the form
\begin{equation}\label{5}
S=i\int d\tau[\bar Z^{\alpha}d Z_{\alpha} + 
l(\tau)\bar Z^{\alpha}Z_{\alpha}],
\end{equation}
where we have added the second term with the Lagrange multiplier $l(\tau)$
which produces the constraint
\begin{equation}\label{6}
\bar Z^{\alpha}Z_{\alpha}=0.
\end{equation}
This constraint implies that $\mu_A$ and $\bar\mu_{\dot A}$ are
determined by eqs. \p{3} and \p{3a}. But, as we shall see below, this 
flat space solution of the twistor constraint \p{6} is not the unique one.
The $AdS_4$ space is also admissible, as well as any space with
a conformally flat metric.

Passing from the action \p{1} to \p{5} we have performed the twistor 
transform, eqs. \p{2}, \p{3} and \p{3a} being the basic twistor relations 
\cite{twistor}.

The action \p{5} is invariant under the conformal $SU(2,2)\sim SO(2,4)$
transformations, since the twistors are in the fundamental representation
of the conformal group. The choice of twistor variables demonstrates 
how conformal symmetry appears in the theory of free massless particles.

We can generalize the action \p{1} to describe a massless particle propagating
in a curved (gravitational) background. For this purpose 
we introduce the vierbein one--form
$e^a(x)=dx^me^a_m(x)$ with the index $a=0,1,2,3$ corresponding 
to local $SO(1,3)$
transformations in the tangent space of the background. Eq. \p{1}
takes the form
\begin{equation}\label{7}
S=\int d\tau \lambda\sigma_a\bar\lambda e^a_m\partial_\tau x^m=
\int\lambda\sigma_a\bar\lambda e^a.
\end{equation}
Note that $\lambda$ still transforms under a spinor representation of
$SO(1,3)\sim SL(2,C)$.

In particular, one can consider an $AdS_4$ space
as a background where the particle propagates. 

Let us mention that a different formulation of
{\em massive} particle mechanics on $AdS_4$ has been considered 
in \cite{kuzenko}.

\subsection{The $AdS_4$ particle}
To consider a particle in the $AdS_4$ background we should specify
the form of $e^a_m(x)$.
A convenient choice of local coordinates and of the form of the 
metric on $AdS_4$ is
\begin{equation}\label{8}
ds^2 =dx^mdx^ne^a_me^b_n\eta_{ab} 
=\left({r\over R}\right)^{2} dx^i \eta_{ij} dx^j 
+ \left(\frac Rr\right)^2 dr^2,
\end{equation}
where $x^m=(x^i,r)$ $(i=0,1,2)$ are coordinates of the $AdS_4$, and
$R$ is the $AdS_4$ radius, whose inverse square is the constant
$AdS_4$ curvature (or the cosmological constant). The coordinates $x^i$
are associated with the three--dimensional  boundary of $AdS_4$
when the radial coordinate $r$ tends to infinity.

  From \p{8} we find that the components of the vierbein one--form 
$e^a=dx^me^a_m$ are
\begin{equation}\label{9}
e^a=dx^me^a_m=\left({r\over R}\right)\delta^a_i dx^i+
\left(\frac Rr\right)\delta^a_3 dr.
\end{equation}

Note that the coordinates $x^i,r$ transform nonlinearly under the action
of the $AdS_4$ isometry group $SO(2,3)$ which is the conformal symmetry
of the boundary $x^i$
\begin{eqnarray}\label{9.1}
\delta x^i&=&a^i_\Pi+a^{ij}_M x_j+a_D x^i+a^i_K x^j x_j-
2( x_ja^j_K)x^i +R^2{a^i_K\over r^2},\\
\delta r&=&(2x_ia^i_K-a_D)r, \nonumber
\end{eqnarray}
where $a^i_\Pi$, $a^{ij}_M$, $a_D$ and $a^j_K$ are parameters of, 
respectively, 
$D=3$ translations, Lorentz rotations, dilatation and conformal boosts,
with $\Pi_i$, $M_{ij}$, $D$ and $K_j$ being the corresponding generators of
the $SO(2,3)$ algebra (i,j=0,1,2) (see Appendix 1). 

We can now substitute \p{9} into \p{7}. The action takes the form
\begin{equation}\label{10}
S=\int d\tau \left[\left({r\over R}\right)
\lambda\sigma_i\bar\lambda\dot x^i+\left(\frac Rr\right)
\lambda\sigma_3\bar\lambda\dot r\right].
\end{equation}

Let us try to carry out the twistor transform of this action in a way
similar to that considered above for the flat target space.
To this end we redefine $\lambda$ as
\begin{equation}\label{11}
\hat\lambda_A=\left({r\over R}\right)^{1\over 2}\lambda_A.
\end{equation}
The action \p{10} takes the form
\begin{equation}\label{11.1}
S=\int d\tau \left[
\hat\lambda\sigma_i\hat{\bar\lambda}\partial_\tau x^i+\left(\frac Rr\right)^2
\hat\lambda\sigma_3\hat{\bar\lambda}\partial_\tau r\right].
\end{equation}
In the limit $r\rightarrow \infty$ one obtains the twistor--like 
massless particle in three--dimensional Minkowski space, i.e. on the 
boundary of $AdS_4$. The complex Weyl spinor $\lambda$ describes a pair of
two--component real D=3 spinors which turn out to be proportional to each 
other on the mass shell. In the limit $r\rightarrow \infty$ 
the action \p{11.1} 
is still invariant under the $D=3$ conformal group $Sp(4)$ =
$O(3,2)/Z_2$ supplemented by $0(2)$ rotations corresponding to the phase
transformations of $\lambda$.

In what follows  we shall, however, keep $r$ finite and 
make the change of variable 
\begin{equation}\label{12}
\hat x^3=-{R^2\over{r}}.
\end{equation}
Then the action \p{11.1} formally becomes the same as eq. \p{1}
in the flat case
\begin{equation}\label{13}
S=\int d\tau \left[
\hat\lambda\sigma_i\hat{\bar\lambda}\partial_\tau x^i+
\hat\lambda\sigma_3\hat{\bar\lambda}\partial_\tau\hat x^3\right]=
\int d\tau\hat\lambda\sigma_m\hat{\bar\lambda}\partial_\tau {\hat x}^m,
\end{equation}
where $\hat x^m=(x^i,\hat x^3)$.
The essential difference is that upon the redefinition \p{11} the 
$SL(2,C)$ spinors $\hat\lambda$ transform {\it nonlinearly} under 
the action of the isometry group $SO(2,3)$ via the radial coordinate $r$.

We can now make the twistor transform of the action \p{13} by
introducing
\begin{equation}\label{14}
\hat{\bar\mu}_{\dot A}=i\hat\lambda^A \hat x_{A\dot A}
\end{equation}
and combining $\hat\lambda$ and $\hat{\bar\mu}$ into the twistor
\begin{equation}\label{15}
{\cal  Z}_{\alpha}=(\hat\lambda_A, \hat{\bar\mu}^{\dot A}).
\end{equation}
The pure twistor form of the action \p{13} is the same as eq. \p{7},
and, hence, it is invariant under the group $SU(2,2)\sim SO(2,4)$
of the conformal transformations acting {\it linearly} on the twistor \p{15}.
The twistor ${\cal Z}_{\alpha}$ is in the fundamental representation of 
$SU(2,2)$.

As it has been explained in detail in \cite{claus2} 
for the particle in $AdS_5$,
the linear conformal $SU(2,2)$ transformations of ${\cal Z}_{\alpha}$
induce nonlinear transformations of the AdS coordinates $x^i$ and $r$ 
when the twistor components are related to $x^i$ and $r$ through  
eqs. \p{11}, \p{12} and \p{14}
\begin{equation}\label{16}
{\cal  Z}_{\alpha}=\left({r\over R}\right)^{1\over 2}\left(\lambda_A, 
-ix^i\sigma_i^{\dot A B}\lambda_B+i{R^2\over r}
\sigma_3^{\dot A B}\lambda_B\right).
\end{equation}
In the case under consideration we thus find the nonlinear conformal  
$SO(2,4)$ transformations of the $AdS_4$ space coordinates, 
with the isometry group
$SO(2,3)$ ( see eq. \p{9.1}) a subgroup of the conformal group.
The conformal transformations of $\hat x^m=x^i,\hat x^3$ (where $\hat x^3$
was defined in \p{12}) are similar to the conformal transformations
of the Minkowski space coordinates. They are
\begin{equation}\label{16.1}
\delta\hat x^m=a^m_\Pi+a^{mn}_M\hat x_n+a_D\hat x^m+a^m_K\hat x^n\hat x_n-
2(\hat x_na^n_K)\hat x^m,
\end{equation}
where $a^m_\Pi$, $a^{mn}_M$, $a_D$ and $a^m_K$ are parameters of, 
respectively, 
$D=4$ translations, Lorentz rotations, dilatation and conformal boosts,
with $\Pi_m$, $M_{mn}$, $D$ and $K_m$ the corresponding generators of
the $SO(2,4)$ algebra $(m,n=0,1,2,3)$ (see Appendix 1).
 
Substituting the expression \p{12} for $\hat x^3$ into \p{16.1}, 
one can deduce the explicit form of the conformal
transformations of the coordinate $r$. Then the $SO(2,3)$ isometry 
transformations \p{9.1} of the $AdS_4$ coordinates are obtained by putting
to zero all parameters in \p{16.1} which carry the index 3, the remaining
ones being $a_D$ and all those with three--dimensional indices $i,j=0,1,2$.

The observation that the $AdS$ spaces allow to be conformally transformed is
implied by the fact that these manifolds are conformally flat. For instance,
the $AdS_4$ metric \p{8} becomes conformally flat upon the 
redefinition of the coordinate $r$ just as in eq. \p{12} (which we made
to perform the twistor transform)
\begin{equation}\label{17}
ds^2 
=\left({R\over{\hat x^3}}\right)^{2} [dx^i \eta_{ij} dx^j +(d\hat x^3)^2]
=\left({R\over{\hat x^3}}\right)^{2}d\hat x^md\hat x^n\eta_{mn}.
\end{equation}

We have thus shown that the twistor constraint \p{6} has two
solutions which correspond to the twistor transform of the flat $D=4$
Minkowski space and of $AdS_4$, both spaces being conformally flat. 
This observation allows us to 
conclude that any other space whose metric is conformally
flat should also arise as a corresponding solution of the 
twistor constraint \p{6}. 

We now turn to the supersymmetrization of the action \p{7}.

\section{Twistor--like $N=1$, $D=4$ superparticles}
The form of the action \p{7} is suitable for a straightforward supersymmetric
generalization. To this end we should consider $e^a$ as a vector component
of the supervielbein one form
\begin{equation}\label{18}
e^I(z)=dz^Me_M^{~I}=(e^a,e^A,\bar e^{\dot A}),
\end{equation}
where $z^M=(x^m,\theta^A,\bar\theta^{\dot A})$ are coordinates which
parametrize a target superspace in which the particle propagates.
$\theta^A$ and its complex conjugate $\bar\theta^{\dot A}$
are Grassmann--odd Weyl spinor coordinates.

\subsection{A superparticle in flat superspace}

In the case of flat target superspace
\begin{equation}\label{19}
e^a= dx^a-i\theta\sigma^a d\bar\theta+id\theta\sigma^a\bar\theta, 
\quad e^A=d\theta^A,
\quad \bar e^{\dot A}=d\bar\theta^{\dot A}.
\end{equation}

Substituting $e^a$ from \p{19} into the action \p{7} we can transform it
into the pure supertwistor action by introducing the supertwistor 
\cite{fer}
\begin{equation}\label{20}
Z_{\cal A}=(\lambda_A, \bar\mu^{\dot A}, \chi)
\end{equation}
and its conjugate
\begin{equation}\label{21}
\bar Z^{\cal A}=(\mu^A, \bar\lambda_{\dot A}, \bar\chi),
\end{equation}
where now
\begin{equation}\label{22}
\bar\mu_{\dot A}=i\lambda^A (x_{A\dot A}
-i\theta_A\bar\theta_{\dot A})
\end{equation}
and
\begin{equation}\label{23}
\chi=\theta^A\lambda_A,\quad \bar\chi=\bar\theta^{\dot A}
\bar\lambda_{\dot A}.
\end{equation}
Upon the supertwistorization the action \p{7} with the supervielbein
\p{19} takes the form similar to eq. \p{5} where the twistor constraint
\p{6} is replaced by the supertwistor constraint
\begin{equation}\label{24}
\bar Z^{\cal A}Z_{\cal A}=
\mu^A\lambda_A-\bar\mu_{\dot A}\bar\lambda^{\dot A}
+2\bar\chi\chi =0.
\end{equation}

For further details on twistor superparticles in flat superspace
we refer the reader to the papers \cite{fer,shir}--\cite{twplyus} 
and proceed with
constructing an action for a superparticle propagating
in the coset superspace ${{OSp(1|4)}\over{SO(1,3)}}$ whose bosonic
subspace is $AdS_4$.

\subsection{The superparticle on ${{OSp(1|4)}\over{SO(1,3)}}$}
To get an explicit form of the particle action on ${{OSp(1|4)}\over{SO(1,3)}}$
we should know an explicit form of the supervielbein \p{18}, 
which is part of the Cartan forms on $OSp(1|4)$. The components
of the latter can be computed using the method of nonlinear realizations 
\cite{ccwz}--\cite{vs}. 

The Cartan forms of the supergroup 
$OSp(1|4)$ and corresponding Cartan forms for the supercoset 
${{OSp(1|4)}\over{SO(1,3)}}$ were calculated in 
\cite{keck}--\cite{ivso}. 
Below we present
simpler expressions for the Cartan forms which allow to write down
a simple form of the $AdS$ superparticle action, since our choice of the
parametrization of the supercoset differs from that used in 
\cite{zumino,ivso}. 

To derive the Cartan forms on ${{OSp(1|4)}\over{SO(1,3)}}$ 
we take the supercoset element in the form
\begin{equation}\label{25}
K(z^M)=B(x)F(\theta)= B(x^m)
e^{i(\theta^A Q_A+\bar\theta_{\dot A}\bar Q^{\dot A})},
\end{equation}
where $B(x^m)$ is the purely bosonic matrix taking its values in the
coset ${{SO(2,3)}\over{SO(1,3)}}$, i.e. it is associated with the bosonic
$AdS_4$ space locally parametrized by coordinates $x^m$. The
Grassmann coordinates
$\theta^A$ and  $\bar\theta^{\dot A}$ extend
$AdS_4$ to the coset superspace, and $Q_A$ and $\bar Q_{\dot A}$
are the odd generators of $OSp(1|4)$ (see Appendix 1).

The Cartan form on ${{OSp(1|4)}\over{SO(1,3)}}$ is
\begin{equation}\label{26}
{1\over i}K^{-1}dK=E^a(z)P_a+E^A(z)Q_A 
+ Q_{\dot A}\bar E^{\dot A}(z)
+\Omega^{ab}(z)M_{ab}.
\end{equation}
It takes values in the $OSp(1|4)$ superalgebra.

The one--forms $E^I=(E^a,E^A,\bar E^{\dot A})$ are the 
supervielbeins and $\Omega^{ab}$ is the $SO(1,3)$ connection on 
the coset superspace.

In the representation \p{25} the Cartan form \p{26}
is
\begin{equation}\label{27}
K^{-1}dK=F^{-1}\left(B^{-1}dB\right)F+F^{-1}dF,
\end{equation}
where the purely bosonic Cartan form $B^{-1}dB$ takes values in
the $SO(2,3)$ algebra and describes a vierbein $e^a(x)$ and 
a spin connection $\omega^{ab}(x)$ on the bosonic $AdS_4$ space
\begin{equation}\label{28}
{1\over i}B^{-1}dB=e^aP_a+\omega^{ab}M_{ab}.
\end{equation}
Depending on the choice of $B(x)$ one can get different forms of 
$e^a(x)$ and $\omega^{ab}(x)$. For instance, the coset element $B(x)$
can be chosen in such a way that $e^a(x)$ in \p{28} is the same as
in \p{9} and the connection $\omega^{ab}(x)$ is
\begin{equation}\label{29}
\omega^{i3}={1\over R}e^i, \quad \omega^{ij}=0,
\end{equation}
(remember that the index 3 corresponds to the radial coordinate $r$
of $AdS_4$).

We can now substitute \p{28} into \p{27} and calculate the explicit
form of the supervielbeins $E^I$ and the superconnection $\Omega^{ab}$,
using a trick, described, for example, in \cite{2bs}, or by the method
presented in Appendix 2. 
In the Majorana spinor representation the expressions for the 
Cartan forms are given in Appendix 1.
In the two--component spinor formalism the supervielbein form
\p{2.41} or \p{2.7} of Appendix 1 can be written as follows
\begin{equation}\label{30}
E^a=
P(\theta^2,\bar\theta^2)\left[e^a(x)-i\Theta\sigma^a D\bar\Theta
+iD\Theta\sigma^a\bar\Theta\right],
\end{equation}
where
\begin{equation}\label{31}
P(\theta^2,\bar\theta^2)=1-{i\over{2R}}(\theta^2-\bar\theta^2)+{1\over{3!R^2}}
\theta^2\bar\theta^2, \qquad \theta^2\equiv \theta^A\theta_A,
\quad \bar\theta^2\equiv\bar\theta_{\dot A}\bar\theta^{\dot A},
\end{equation}
\begin{equation}\label{31.1}
\Theta=\left({{1+{i\over {3!R}}(\theta^2-\bar\theta^2)}
\over{P(\theta^2,\bar\theta^2)} }\right)^{1\over 2}\theta,
\end{equation}
$$
{D}=d+\omega^{bc}\sigma_{bc}, \quad 
\sigma^{ab}={1\over 4}(\sigma^a\bar\sigma^b-\sigma^b\bar\sigma^a).
$$

To get the action for the superparticle in the super--AdS background
we should simply substitute \p{30} into \p{7}. 
\begin{equation}\label{32}
S=\int\lambda\sigma_a\bar\lambda 
P(\theta^2,\bar\theta^2)\left[e^a-i\Theta\sigma^a D\bar\Theta
+iD\Theta\sigma^a\bar\Theta\right]
\end{equation}
The polynomial 
$P(\theta^2,\bar\theta^2)$ can be absorbed by properly rescaled $\lambda$
and $\bar\lambda$, namely, $\Lambda=\sqrt{P(\theta^2,\bar\theta^2)}\lambda$.
Then the action takes an even simpler form which is quadratic in fermions
\begin{eqnarray}\label{32.1}
S&=&\int\Lambda\sigma_a\bar\Lambda \left[
e^a-i\Theta\sigma^a {{D}}\bar\Theta
+i{D}\Theta\sigma^a\bar\Theta\right] \\
&=&\int\Lambda\sigma_a\bar\Lambda \left[
e^a-i\Theta\sigma^a d\bar\Theta
+id\Theta\sigma^a\bar\Theta
+2i\omega^{bc}(x)\Theta\sigma^a\bar\sigma_{bc}\bar\Theta\right] \nonumber.
\end{eqnarray}
If in \p{32.1} there were no term containing the spin connection
$\omega^{bc}$ 
the action \p{32.1} could be completely
supertwistorized in the same way as we have done in the case of
the $AdS_4$ particle and of the superparticle in flat superspace.
However, the term with  $\omega^{bc}$ does
not allow one to perform the complete supertwistorization of \p{32.1}
in terms of {\it free} supertwistors, 
at least in a straightforward way.

Using the notion of Killing spinors on AdS spaces one can replace in
\p{32.1} the covariant differential $D$ with the ordinary one.
To this end it is convenient to switch to the four--component Majorana spinor
formalism
$$
\Lambda^{\alpha}=(\lambda_A,\bar\lambda^{\dot A}), \quad
\Theta^{\alpha}=(\Theta_A,\bar\Theta^{\dot A}).
$$

By definition (see, for instance, \cite{killing}) 
$AdS$ Killing spinors satisfy the condition
\begin{eqnarray}\label{33}
{\cal D}K^{\alpha}_{~\beta}C^\beta&=&(DK^{\alpha}_{~\beta}
+{1\over{2R}}(e^a\gamma_a)^\alpha_{~\gamma}K^{\gamma}_{~\beta})C^\beta \nn
&=&
(dK^{\alpha}_{~\beta}
+{1\over 2}(\omega^{ab}\gamma_{ab})^\alpha_{~\gamma}K^{\gamma}_{~\beta}
+{1\over{2R}}(e^a\gamma_a)^\alpha_{~\gamma}K^{\gamma}_{~\beta})C^\beta=0,
\end{eqnarray}
where $K^{\alpha}_{~\beta}(x)$ is a bosonic Killing spinor matrix and 
$C^\beta$ is an arbitrary constant spinor.
If in \p{32.2} we replace $\Theta$ 
with $\Theta=K(x)\Theta_K$ (where $\Theta_K\equiv K^{-1}\Theta$) \cite{renata}
the action \p{32.1} 
takes the form
$$
S=\int\bar\Lambda\gamma_a\Lambda \left[
e^a(1+{i\over{2R}}\bar\Theta\Theta)-i\bar\Theta\gamma^a Kd\Theta_K\right],
$$
or (upon an appropriate rescaling of $\Lambda$ and $\Theta$)
\begin{equation}\label{32.2}
S=\int\bar\Lambda\gamma_a\Lambda \left[
e^a-i\bar\Theta\gamma^a Kd\Theta_K\right].
\end{equation}
Note that in \p{32.2} the variable $\Theta_K$ is regarded as independent, 
while $\Theta=K(x)\Theta_K$ is composed from $\Theta_K$ and the Killing
matrix $K(x)$ whose exact dependence on the $AdS_4$ coordinates $x^m$ can be
found by solving the Killing spinor equation \p{33} \cite{killing}. 
Taking this into account, the term $i\bar\Theta\gamma^a Kd\Theta_K$ in \p{32}
can be rewritten as 
$i\bar\Theta_K\gamma^{\hat b\hat c}d\Theta_KK^a_{\hat b\hat c}(x)$, 
where $K^a_{\hat b\hat c}(x)$ are $SO(2,3)$ Killing vectors on $AdS_4$
($\hat b,\hat c=0,1,2,3,4$, $\gamma^4=${\bf 1}). Then the supervielbein
$$
E^a=e^a-i\bar\Theta_K\gamma^{\hat b\hat c}d\Theta_KK^a_{\hat b\hat c}
$$
takes the form similar to one of the parametrizations 
considered in \cite{zumino}.

It would be interesting to understand whether the $AdS$ superparticle action
in any of its forms can be completely supertwistorized. If it is possible,
the $AdS$ superparticle model would acquire the manifest superconformal
$SU(2,2|1)$ symmetry.
In any case, the use of commuting spinors whose bilinears replace the
conventional particle momentum and the suitable choice of the parametrization
of the supercoset space ${{OSp(1|4)}\over{SO(1,3)}}$ allow one to get 
a simple form of the
action for a superparticle propagating in the AdS superbackground,
which is bilinear in fermionic variables.

\section{The superparticle on $OSp(1|4)$}
We now turn to the construction of the action for a superparticle propagating
on the supergroup manifold $OSp(1|4)$ locally parametrized by the
supercoset ${{OSp(1|4)}\over{SO(1,3)}}$ coordinates $x^m$ and $\theta$, and by
six $SO(1,3)$ group coordinates $y^{mn}=-y^{mn}$. This model is intended to
produce, upon an appropriate contraction, 
the superparticles in flat superspace and on 
super--$AdS_4$ considered above, as well as the superparticle with tensorial
central charges \cite{1/4,bls}.

By analogy with eqs. \p{7} and \p{32}, to construct the $OSp(1|4)$ 
superparticle Lagrangian 
we take the pullback onto the particle worldline of the
even  Cartan superforms $E^a_{OSp}$ and
$\Omega^{ab}_{OSp}$ given in Appendix 1 (eq. \p{A3}). 
These forms comprise the bosonic $SO(2,3)$ part of the supervielbein 
on $OSp(1|4)$.    
We contract them 
with commuting spinor bilinears $\bar\lambda\gamma_a\lambda$ and
${1\over 2}\bar\lambda\gamma_{ab}\lambda$. The $OSp(1|4)$ 
superparticle action is
\begin{equation}\label{34}
S_{OSp}={1\over 2}\int\left
[E^b(x,\theta)u^{~a}_b(y)\bar\lambda\gamma_a\lambda+
{1\over 2}\left(\Omega^{cd}(x,\theta)u^{~a}_cu^{~a}_d+(u^{-1}du)^{ab}\right)
\bar\lambda\gamma_{ab}\lambda\right].
\end{equation}
Using the defining relations for the $SO(1,3)$ matrices $u^{~a}_b$
and $v^{~\alpha}_\beta$ \p{vu} we can make the redefinition
\begin{equation}\label{35}
u^{~a}_b(y)\bar\lambda\gamma_a\lambda=\bar{\hat\lambda}\gamma_b\hat\lambda,
\quad {\rm where}\quad \hat\lambda^\alpha=\lambda^\beta v^{~\alpha}_\beta.
\end{equation}
Then  $u^{~a}_b(y)$ remains only in one term of the action \p{34}, and
the latter takes the form
\begin{equation}\label{36}
S_{OSp}={1\over 2}\int E^a(x,\theta)\bar{\hat\lambda}\gamma_a\hat\lambda+
{1\over 4}\int [\Omega^{ab}(x,\theta)+(duu^{-1})^{ab}]
\bar{\hat\lambda}\gamma_{ab}\hat\lambda.
\end{equation}
We observe that the first integral in \p{36} is nothing but the action \p{32}
for the superparticle on the coset superspace ${{OSp(1|4)}\over{SO(1,3)}}$,
and the second term contains the spin connection of  
${{OSp(1|4)}\over{SO(1,3)}}$ extended by the $SO(1,3)$ Cartan form $duu^{-1}$.
In eq. \p{34} the dependence of the action on the $SO(1,3)$ group manifold
coordinates $y^{mn}$ remains only in $duu^{-1}$.

Since by an appropriate choice of Grassmann coordinates
the Cartan forms $E^a(x,\theta)$ and  $\Omega^{ab}(x,\theta)$ can be made
quadratic in $\theta$ (see eqs. \p{2.7} and \p{2.8} of Appendix 1) we 
see that the $OSp(1|4)$ action \p{36} is {\it quadratic} in fermions.

If we drop the second integral of \p{36} we get the action for the
superparticle considered in Subsection 3.2, and if we then take the limit
when the $AdS_4$ radius goes to infinity, the action further reduces to
the superparticle action in flat $N=1$, $D=4$ superspace. 

Another way of truncating the action \p{36} is to perform the following
contraction of the $OSp(1|4)$ superalgebra \p{2.1}. Let us in \p{2.1}
rescale the generators $M_{ab}$ of $SO(1,3)$ as follows
\begin{equation}\label{37}
M_{ab}=RZ_{ab},
\end{equation}
and consider the limit when $R\rightarrow \infty$.
Then the generators $Z_{ab}$ become tensorial central charges which commute
with all other generators, and the anticommutator of the supercharges becomes
\begin{equation}\label{38}
\left\{{Q}_{\alpha}, Q_{\beta}\right\}
=  - 2(C\gamma^{a})_{\alpha\beta}P_{a}
  + \left(C\gamma^{ab}\right)_{\alpha\beta}
Z_{ab}.
\end{equation}
The $SO(1,3)$ coordinates $y^{mn}$ become central charge coordinates.

In the limit $R\rightarrow \infty$ the supervielbein $E^a(x,\theta)$ 
reduces to the `flat' one--form \p{19}
\begin{equation}\label{39.1}
E^a_Z=dx^a-i\bar\theta\gamma^a d\theta
\end{equation}
 and the superconnection 
$\Omega^{ab}_Z=R(\Omega^{ab}+(duu^{-1})^{ab})$ becomes
\begin{equation}\label{39}
\Omega^{ab}_Z=dy^{ab}+{i\over 2}\bar\theta\gamma^{ab}d\theta.
\end{equation}

Substituting \p{39.1} and \p{39} into \p{36} 
we get the action for a particle with tensorial central charges
\cite{1/4,bls}. 
The quantization of this  superparticle model has shown to produce
a infinite tower of free massless states with arbitrary integer and 
half integer spin, with the spin degrees of freedom associated with 
the central charge coordinates $y^{mn}$. For a detailed analysis of the model 
we refer the reader to \cite{1/4,bls}.

Since the higher--spin fields can interact if they live not in Minkowski space
but in an anti--de--Sitter space \cite{fv}, it seems of interest to study the
possibility of generalizing the $OSp(1|4)$ superparticle model based on the
action \p{36} to include interactions, and then to perform its quantization 
to check whether such a model can be considered as a classical counterpart
of the theory of interacting higher--spin fields.

To conclude this section we demonstrate that $OSp(1|4)$ covariant momenta
associated with the $OSp(1|4)$ coordinates $x^m,y^{mn}$ and $\theta^\alpha$
generate the $OSp(1|4)$ superalgebra \footnote{In \cite{abs} similar 
covariant momenta were used to make the Hamiltonian analysis and the 
quantization of superparticles propagating in harmonic superspaces.}.

Let us rewrite the action \p{36} as follows
\begin{equation}\label{40.2}
S=\int d\tau\Lambda_{I}E^{I}_{M}(x,\theta,z)\partial_\tau X^{M},
\end{equation}
where $\Lambda_{I}$ stand for the bilinear combinations of the spinor 
$\hat\lambda$
\begin{equation}\label{40.1}
\Lambda_{I}=\left({1\over 2}\hat\lambda\gamma_a\hat\lambda,~
{1\over 4}\hat\lambda\gamma_{ab}\hat\lambda\right),
\end{equation}
the index $I$ stands for vector $a$ and tensor $ab$ indices, and
$X^{M}\equiv(x^m,z^{mn},\theta^\alpha)$. $E^{I}_{M}(x,\theta,z)$ are the
$OSp(1|4)$ Cartan form components $E^a$ and $\Omega^{ab}$, 
which correspond to the bosonic generators
$P_a$ and $M_{ab}$ of $OSp(1|4)$ (see Appendix 1).

 From \p{40.2} we get the canonical momenta conjugate to 
$X^{M}\equiv(x^m,z^{mn},\theta^\alpha)$ as
\begin{equation}\label{40.3}
{{\delta S}\over{\delta(\partial_\tau X^M)}}=P_{M}=\Lambda_I E^{I}_{M}.
\end{equation}
Multiplying \p{40.3} by the matrix $E_{\hat I}^{ M}$ inverse to 
$E^{\hat I}_{ M}$
(where $\hat I=(I,\alpha)$) we obtain  
$OSp(1|4)$ covariant momenta 
$P_{\hat I}=E_{\hat I}^{M}P_{M}=(P_I,P_\alpha)$ such that
\begin{equation}\label{40.4}
\Lambda_I=P_I\equiv E_{ I}^{M}(X)P_{M},\qquad 
P_\alpha=E_{\alpha}^{M}(X)P_{M}=0.
\end{equation}
Eqs. \p{40.1} and \p{40.4} imply
that the expressions for the momenta are constraints 
on the superparticle phase space variables. For instance, 
the momentum components $P_\alpha$ of the Grassmann variable 
$\theta^\alpha$ are zero. These are Grassmann constraints on the dynamics
of the $OSp(1|4)$ superparticle, which include first--class constraints
generating the $\kappa$--symmetry of the $OSp(1|4)$ superparticle.

It is well known that, as $N=1$, $D=4$ superparticles in an arbitrary 
supergravity background do, 
the $AdS$ superparticle possesses two--parameter  
local fermionic $\kappa$--symmetry, which means that such superparticles
preserve half the supersymmetry of a target--space vacuum.

In contrast to this, as we shall prove in the next section, 
the $OSp(1|4)$ superparticle 
possesses 3 $\kappa$--symmetries and, in general,  
the superparticle propagating on the $OSp(1|2n)$ supergroup manifold 
has $(2n-1)$ $\kappa$-symmetries and thus describes  BPS 
states with only one broken supersymmetry. 

In \cite{1/4} the superparticle models with such a symmetry 
property have been obtained in flat 
superspaces with additional tensorial central charge coordinates. 
Here we observe that this unusual feature is also inherent to 
superparticles propagating in more complicated superspaces. 

Because of the Maurer--Cartan equations $(dE-iE\wedge E=0)$
for the Cartan forms $E^{\hat I}_{M}$  
the generalized momenta form, under the Poisson brackets,
the $OSp(1|4)$ superalgebra, which can be quantized by taking
an appropriate ordering of $X$ and $P$ in the definition of \p{40.4}:
\begin{equation}\label{40.5}
[P_{\hat I},P_{\hat J} \} =C_{\hat I\hat J}^{~~~\hat K}P_{\hat K},
\end{equation}
where $C_{\hat I\hat J}^{~~~\hat K}$ are $OSp(1|4)$ superalgebra 
structure constants (see eq. \p{2.1} of Appendix 1).

 From \p{40.4} and \p{40.5} we see that upon the transition to 
Dirac brackets 
the spinor bilinears $\Lambda_I$ \p{40.1} become generators of the 
$Sp(4)\sim SO(2,3)$ subalgebra of $OSp(1|4)$.

 From this analysis we conclude that the commutation properties of the 
superparticle covariant momenta reflect the structure 
of the global symmetries of the $OSp(1|4)$ superparticle action.
To quantize the model one should consider the $OSp(1|4)$ coordinates and
momenta as `generalized' canonical variables, with commutation
relations defind by the graded $OSp(1|4)$ superalgebra \p{40.5}.

The detailed study of the model based on the action \p{36} is in progress.

\section{The superparticle on $OSp(1|2n)$ as a dynamical 
model for unusual BPS states}

We now generalize the $OSp(1|4)$ superparticle action \p{34}, \p{36} or
\p{40.2} to the case of the supermanifold $OSp(1|2n)$ whose parametrization
we choose to be of the form (see Appendix 2 for the details on the 
$OSp(1|2n)$ superalgebra)
\begin{equation}\label{41}
{\cal G}(y,\theta)=B(y)F(\theta)=B(y)e^{i\theta^\alpha Q_\alpha},
\end{equation}
where $y^{\alpha\beta}=y^{\beta\alpha}$ are coordinates of the
$Sp(2n)$ subgroup generated by symmetric operators
$M_{\alpha\beta}=M_{\b\a}$, and
whose element is denoted as $B(y)$;
$\theta^\alpha$ are Grassmann coordinates and $Q_\alpha$ are Grassmann
generators of $OSp(1|2n)$ transforming under the fundamental 
representation of $Sp(2n)$, which we call the spinor representation, 
$(\alpha,\beta=1,...,2n)$.

The $OSp(1|2n)$ Cartan forms are
\begin{equation}\label{Cf}
{1\over i}{\cal G}^{-1} (y , \th ) d {\cal G} (y , \th ) 
\equiv
{1\over i} \left[ F^{-1} (B^{-1} dB) F + F^{-1} dF\right]  
= {E}^\a Q_\a +  {1\over 2} \Om^{\a\b} M_{\a\b}.
\end{equation}

To have the connection with the $OSp(1|4)$ case discussed in
Section 4 and Appendix 1 we note that for $n=2$ $M_{\a\b}$
can be written in terms of $SO(1,3)$ covariant generators 
$P_a$ and $M_{ab}$
as follows
\begin{equation}\label{PaM}
M_{\a\b}=-2(C\gamma^a)_{\a\b}P_a+{1\over R}(C\gamma^{ab})_{\a\b}M_{ab}.
\end{equation}
Then the $OSp(1|4)$ Cartan forms presented in \p{A3} are related to
$\Om^{\a\b}$ in \p{Cf} in the following way
\begin{equation}\label{42}
E^a_{OSp}=-(C\gamma^a)_{\alpha\beta}\Om^{\a\b},
\quad \Om^{ab}_{OSp}= {1\over {2R}}(C\gamma^{ab})_{\a\b}\Om^{\a\b}.
\end{equation}
The matrix $C_{\a\b}$ plays the role of the $OSp(1|2n)$ invariant metric.

The $OSp(1|2n)$ Cartan forms \p{Cf} computed in Appendix 2 have
the form
\begin{equation}\label{Ea(D)}
{E}^{\a} = {\cal D}\th^\a +i{\cal D}\th^{(\a } \th^{\b )} \th_\b 
P_1(\theta\theta), 
\end{equation}
\begin{equation}\label{Om(D)}
\Om^{\a\b} = 
\om^{\a\b}(y) 
+ i\th^{(\a }  {\cal D}\th^{ \b )}P_2(\theta\theta),
\end{equation}
where $\om^{\a\b}(y)$ are $Sp(2n)$ Cartan forms, $P_1(\theta\theta)$
and $P_2(\theta\theta)$
are polynomials in $\theta^\a C_{\a\b}\theta^\b$ (see \p{poly1} and \p{poly2} 
of Appendix 2),
and ${\cal D}$ is the $Sp(2n)$ covariant derivative
\begin{equation}\label{Dth}
{\cal D}\th^{\a}= d\th^{\a} + {\alpha\over 2} \om^\a_{~\b}(y)\th^\b ,
\end{equation}
where $\alpha$ is a dimensional constant factor in the $OSp(1|2n)$
superalgebra \p{alg}, which in the $OSp(1|4)$ case \p{2.1} is
$\alpha={{4}\over R}$. 

The form of eq. \p{Om(D)} prompts us that the polynomial $P_{2}$ can
be hidden into rescaled $\Theta=\sqrt{P_{2}}\theta$, then for 
$\Om^{\a\b}$ we get the simple expression
\begin{equation}\label{simple}
\Om^{\a\b} = 
\om^{\a\b}(y) 
+i \Theta^{(\a }{\cal D} \Theta^{ \b )}.
\end{equation}

The action for a superparticle moving on $OSp(1|2n)$, 
which generalizes \p{36}, has the form 
 \begin{equation}\label{Sn}
S= {1 \over 2} \int \l_\a \l_\b \Om^{\a\b} \equiv 
{1 \over 2} \int d\tau \l_\a \l_\b \Om_\tau^{\a\b} 
\end{equation}
where $\l_\a$ is an auxiliary bosonic $Sp(2n)$ `spinor' variable, 
and
$ \Om^{\a\b}=  d\tau \Om_\tau^{\a\b}$ is the  pullback of the 
even Cartan form \p{Om(D)} or \p{simple} on the superparticle
worldline. 

Let us now analyse the $\kappa$--symmetry properties of the action \p{Sn}
by considering its general variation. 
A simple way to vary the action \p{Sn} with respect to
$OSp(1|2n)$ coordinates $X^M=(y^{\a\b},\theta^{\a})$ 
and the auxiliary variable $\lambda$, 
is to use   
Maurer--Cartan equations (integrability conditions for Eq. \p{Cf})
$d({\cal G}^{-1}d{\cal G})= {\cal G}^{-1}d{\cal G} \wedge 
{\cal G}^{-1}d{\cal G}$ which imply
 \begin{equation}\label{MC1}
dE^\a + {\alpha\over 2} E^\b \wedge \Om_\b^{~\a} = 0, \qquad 
\end{equation}
 \begin{equation}\label{MC2}
d\Om^{\a\b} + {\alpha\over 2} \Om^{\a \g} \wedge \Om_\g^{~\b} = 
-{i} E^\a \wedge E^\b , \qquad 
\end{equation}
and the expression for the $X^M$--variation of the differential forms
 \begin{equation}\label{dd}
\d \Om = i_\d d\Om + di_\d \Om  \quad i_\d \Om\equiv \d X^M\Om_M. 
\end{equation} 

Modulo a boundary term the variation of the action \p{Sn} obtained in
this way takes the form
  \begin{equation}\label{dSn}
S=  \int \d \l_\a \Om^{\a\b} \l_\b -  
\int_{{\cal M}^1} {\cal D} \l_{\a }~ i_\d\Om^{\a\b} \l_\b - 
{i\over 2} \int (E^\a \l_\a ) 
(i_\d E^\b) \l_\b ,
\end{equation}
where the basis in the space of variations 
is choosen to be 
$i_\d \Om^{\a\b}$ and 
$i_\d E^{\a}$ instead of more conventional $\d y^{\a\b}$ and $\d \th^\a$.

Note that $i_\d E^{\a}$ corresponds to the variation of the action
with respect to Grassmann coordinates $\theta^\a$.
Putting $\d \l_\a =0, ~i_\d \Om^{\a\b}=0$ we thus observe that only
one of the $2n$ independent fermionic variations, namely
 $i_\d E^\a \l_\a$, effects the variation of the action. 
This implies that other
$2n-1$ fermionic variations are fermionic $\kappa$--symmetries 
of the dynamical system described by the action \p{Sn}. 
The $\kappa$--symmetry transformations are defined in such a way
that $i_\d E^\a \l_\a$ vanishes (cf. \cite{1/4,bls})
\begin{equation}\label{kappan}
i_\d \Om^{\a\b} =0, \qquad \d \l_\a = 0, \qquad i_\d E^\a = 
\k^I \mu^\a_I, \quad I=1,\ldots (2n-1)
\end{equation}
where the $\mu^\a_I$ are $2n-1$ $Sp(2n)$ spinors orthogonal to $\l_\a$ 
\begin{equation}\label{mun}
\mu^\a_I \l_\a = 0, \qquad I=1, \ldots (2n-1). 
\end{equation}

Thus, we conclude that an unusual property
of a twistor--like superparticle with tensorial central charge coordinates 
\cite{1/4} to preserve all but one target--space supersymmetries
is inherent to the superparticle model on the $OSp(1|2n)$ 
supergroup manifold as well. 

When the explicit expressions \p{Ea(D)} and  \p{Om(D)} 
for the Cartan forms on $OSp(1|2n)$ are obtained,
one straightforwardly gets the explicit expressions also for the Cartan forms
on any coset superspace $OSp(1|2n)/H$, where $H$ is a bosonic subgroup of
$OSp(1|2n)$. These expressions are the same as \p{Ea(D)} and  \p{Om(D)} but
with $\omega^{\a\b}$ depending only on the bosonic coordinates 
of the supercoset (see also eqs. \p{Asimple} and \p{alsimple} of Appendix 2).

Using the $OSp(1|2n)/H$ Cartan forms one can construct various types of 
actions for superparticles and superbranes propagating on the corresponding 
coset supermanifolds.

\section{Conclusion}

By taking a suitable parametrization of the supergroup manifold  
$OSp(1|2n)$ we have found a simple form of the $OSp(1|2n)$ Cartan superforms
such that the ones which take values in the bosonic
subalgebra $Sp(2n)$ of $OSp(1|2n)$ are quadratic in 
Grassmann coordinates.

We have used these Cartan forms to construct simple 
twistor--like actions (which are quadratic in fermions)
for describing superparticles propagating on the coset superspace 
$OSp(1|4)/SO(1,3)$, on the supergroup manifold $OSp(1|4)$, and, in general,
on $OSp(1|2n)$  supermanifolds.  

The $OSp(1|4)$ superparticle model has been shown to produce (upon a
truncation) either the standard massless D=4 superparticle or the
generalized massless D=4 superparticle with
tensorial central charges \cite{1/4,bls} whose quantization gives rise
to massless free fields of arbitrary (half)integer spin.

We have also shown that the massless particle on $AdS_4=SO(2,3)/SO(1,3)$
can be described (with a particular choice of twistor variables) as a free
$D=4$ twistor particle.

A direction of further study can be to analyse the 
$OSp(1|4)$ superparticle model in detail and to look for its role
as a classical counterpart in the theory of interacting
higher--spin fields \cite{fv}--\cite{misha3} requiring a finite
AdS radius.

Another interesting problem is to generalize the results of this
paper to the case of superstrings and superbranes propagating in 
AdS superbackgrounds with the aim to find a simple form of superbrane actions 
on AdS. The simple fermionic structure of $OSp(1|32)$ and $OSp(1|64)$ 
Cartan forms, which we obtained, may be helpful in making a progress in 
this direction.

\bigskip
\noindent
{\bf Acknowledgements}
I.B. acknowledges the financial support from the Austrian Science
Foundation 
under the project M472-TPH, and D.S. acknowledges the financial
support from the Alexander von Humboldt Foundation. 
Work of I.B. and D.S. was also partially supported by 
the INTAS Grant  N96--308.
I.B. is grateful to Prof. M. Virasoro and Prof. S. Ranjbar-Daemi for the
hospitality at the ICTP on the final stage of this work. 
J. L. would like to thank Prof. J. Wess and Max--Planck--Institute 
f\"ur Physik in Munich for hospitality and financial support.

\section*{Appendix 1}
\def\theequation{A.\arabic{equation}}
\setcounter{equation}0

We use the `almost plus' signature $(-,+,\cdots,+)$
of the Minkowski metric $\eta^{ab}$ $(a,b=0,1,2,3)$.

\subsection*{The $OSp(1|4)$ superalgebra}
\begin{eqnarray}\label{2.1}
 -i[M_{ab},M_{cd}]&=& 
\eta_{ad}M_{bc} + \eta_{bc}M_{ad} 
- \eta_{ac}M_{bd} -\eta_{bd}M_{ac}
\label{2.1aa}
\\ \cr
 -i[M_{ab},P_{c}]&= & \eta_{bc}P_{a}-\eta_{ac}P_{b}
 \label{2.1b}
\\ \cr 
 [P_{a},P_{b}] &= & {i\over R^{2}} M_{ab}
 \label{2.1c}
\\ \cr 
 \left\{{Q}_{\alpha}, Q_{\beta}\right\}
&= & - 2(C\gamma^{a})_{\alpha\beta}P_{a}
  + {1\over R} \left(C\gamma^{ab}\right)_{\alpha\beta}
M_{ab}
\label{2.1d}
\cr
\cr  
 [M_{ab},Q_{\alpha}] &= &- {i\over 2} Q_{\beta}
\left(\gamma_{ab}\right)^{\beta}_{\ \alpha} 
\quad \gamma_{ab}={1\over 2}(\gamma_a\gamma_b-\gamma_b\gamma_a),
\label{2.1e}
\\ \cr
 [P_{a},Q_{\alpha}] & =&-
 {i\over 2R} Q_{\beta}
\left(\gamma_{a}\right)^{\beta}_{\ \alpha}
\label{2.1f}
\end{eqnarray}
The generators $M_{ab}$ form the $SO(1,3)$ subalgebra \p{2.1aa},
and $M_{ab}$ and $P_a$ form the $SO(2,3)$ subalgebra of $OSp(1|4)$.
$Q_\alpha$ are four Majorana spinor generators of $OSp(1|4)$.
The parameter $R$ is the $AdS_4$ radius, and $C_{\alpha\beta}$ is
the charge conjugation matrix such that
$$
\gamma^a_{\alpha\beta}=\gamma^a_{\beta\alpha}\equiv 
C_{\alpha\gamma}(\gamma^a)^\gamma_{~\beta}.
$$

The parameters  $a^i_\Pi$, $a^{ij}_M$, $a_D$ and $a^i_K$
\p{9.1} of $SO(2,3)$ acting as the conformal transformations
on the boundary of $AdS_4$ (associated with the coordinates $x^i$) correspond
to the following linear combinations of $M_{ab}$ and $P_a$:\\
three--dimensional translations
$$
a^i_\Pi \quad \rightarrow \quad \Pi_i=P_i-M_{i3} \quad i=0,1,2\,;
$$
$$
[\Pi_i,\Pi_j]=0
$$
$SO(1,2)$--rotations
$$
a^{ij}_M \quad \rightarrow \quad M_{ij},
$$
dilatation
$$
a_D\quad \rightarrow \quad D=P_3,
$$
special conformal transformations (conformal boosts)
$$
a^i_K \quad \rightarrow \quad K_i=P_i+M_{i3}.
$$

Note that the $SO(2,4)$ algebra has the same structure as $SO(2,3)$ in 
(A.1)--(A.3) but with indices $a,b, \cdots$ running from 0 till 4.

\subsection*{The $OSp(1|4)$ Cartan forms}
We choose the parametrization of an $OSp(1|4)$ group element $G(x,\theta,y)$
as follows
\begin{equation}\label{A2}
G=K(x,\theta)U(y), \quad K(x,\theta)=B(x)e^{i\theta Q},
\end{equation} 
where
$
K(x,\theta)=B(x)e^{i\theta Q}
$
is a group element  corresponding to 
the coset superspace  ${{OSp(1|4)}\over{SO(1,3)}}$,
$B(x)$ is a group element corresponding to the bosonic 
$AdS_4={{SO(2,3)}\over{SO(1,3)}}$ and $U(y)$ is an element of 
$SO(1,3)$ generated by $M_{ab}$ with the antisymmetric $y^{ab}$ being
six parameters of the  $SO(1,3)$ transformations. We do not need to
specify the representation of $B(x)$ and $U(y)$.

The $OSp(1|4)$ Cartan forms $G^{-1}dG=E^a_{OSp}P_a+\Omega^{ab}_{OSp}M_{ab}+
E^{\alpha}_{OSp}Q_{\alpha}$ are
\begin{eqnarray}\label{A3} 
E^a_{OSp}&=&E^b(x,\theta)u^{~a}_b(y),\nn
\Omega^{ab}_{OSp}&=&\Omega^{cd}(x,\theta)u^{~a}_cu^{~a}_d+(u^{-1}du)^{ab},\nn
E^{\alpha}_{OSp}&=&E^{\beta}(x,\theta)v_{\beta}^{~\alpha}(y),
\end{eqnarray}
where $u^{~a}_b(y)$ and $v_{\beta}^{~\alpha}(y)$ are  matrices of,
respectively, the vector and the spinor representation of $SO(1,3)$. 
They are defined by the relations
\begin{equation}\label{VU}
u^{~a}_b(y)P_a=U^{-1}P_bU(y), \quad v_{\beta}^{~\alpha}(y)Q_\alpha=
U^{-1}Q_\beta U(y),
\end{equation}
and are related to each other
by the standard expression\begin{equation}\label{vu}
\gamma_au^{~a}_b(y)=v(y)\gamma_bv(y).
\end{equation}
$E^a(x,\theta)$, $\Omega^{ab}(x,\theta)$ and $E^{\alpha}(x,\theta)$
are Cartan forms $K^{-1}dK$ 
corresponding to the coset superspace ${{OSp(1|4)}\over{SO(1,3)}}$.

\subsection*{The  ${{OSp(1|4)}\over{SO(1,3)}}$ supervielbeins and
spin connection}
The spinorial supervielbein is
\begin{equation}\label{2.2}
E^\alpha={\cal D}\theta^\alpha 
-{i\over {3!R}}\bar\theta\gamma^a{\cal D}\theta(\gamma_a\theta)^\alpha+
{i\over {2\cdot 3!R}}
\bar\theta\gamma^{ab}{\cal D}\theta(\gamma_{ab}\theta)^\alpha
-{2\over{5!R^2}}{\cal D}\theta^\alpha 
(\bar\theta\theta)^2,
\end{equation}
or, by using the Fierz identity
\begin{equation}\label{fierz}
C_{\alpha(\beta}C_{\gamma)\delta}=
{1\over 4}\gamma^a_{\beta\gamma}(\gamma_a)_{\alpha\delta}-
{1\over 8}\gamma^{ab}_{\beta\gamma}(\gamma_{ab})_{\alpha\delta},
\end{equation}
\begin{equation}\label{2.2f}
E^\alpha={\cal D}\theta^\alpha(1+{i\over{ 3R}}\bar\theta\theta
-{2\over{5!R^2}}(\bar\theta\theta)^2)
-{i\over{ 3R}}\bar\theta{\cal D}\theta\theta^\alpha.
\end{equation}
where ${\cal D}$ is a covariant differential on the bosonic $AdS_4$ space
defined as
\begin{equation}\label{2.3}
{\cal D}=d+{1\over 2}\omega^{ab}(x)\gamma_{ab}+{1\over{2R}}e^a(x)\gamma_a
\equiv D+{1\over{2R}}e^a\gamma_a.
\end{equation}
Note that the $AdS_4$ Killing spinors \p{33} are defined to be 
covariantly constant with respect to ${\cal D}$, i.e. ${\cal D}K=0$.

The vector supervielbein is 
$$
E^a=e^a(x)-i\bar\theta\gamma^a{\cal D}\theta
-{1\over {2\cdot 3!R}}\bar\theta\gamma^a{\cal D}\theta(\bar\theta\theta)
+{1\over{4!R}}\bar\theta\gamma^{bc}{\cal D}\theta(\bar\theta\gamma^a
\gamma_{bc}\theta),
$$
or (upon applying the Fierz identity \p{fierz})
\begin{equation}\label{2.4}
E^a=e^a(x)-i\theta\gamma^a{\cal D}\theta(
1+{i\over {3!R}}\bar\theta\theta).
\end{equation}
Eq. \p{2.4} can be further rewritten as
\begin{equation}\label{2.41}
E^a=e^a(x)\left(1-{i\over{2R}}\bar\theta\theta-{1\over {2\cdot 3!R^2}}
(\bar\theta\theta)^2\right)-i\theta\gamma^a{D}\theta\left(1
+{i\over {3!R}}\bar\theta\theta\right),
\end{equation}
where $D=d+{1\over 2}\omega^{ab}(x)\gamma_{ab}$.

And the $SO(1,3)$ connection is
$$
\Omega^{ab}=\omega^{ab}(x)+{i\over{2R}}\bar\theta\gamma^{ab}{\cal D}\theta
+{1\over{4!R^2}}(\bar\theta\gamma^c{\cal D}\theta)(\bar\theta\gamma^{ab}
\gamma_c\theta)-{1\over{2\cdot 4!R^2}}(\bar\theta\gamma^{cd}{\cal D}\theta)
(\bar\theta\gamma^{ab}\gamma_{cd}\theta)
$$
\begin{equation}\label{2.5}
=\omega^{ab}(x)+{i\over{2R}}\bar\theta\gamma^{ab}{\cal D}\theta
\left(1+{i\over{3!R}}\bar\theta\theta\right),
\end{equation}
where $e^a(x)$ and $\omega^{ab}(x)$ are the vierbein 
and the spin connection on $AdS_4$.

Note that in \p{2.4} and
\p{2.5}  we can make the following change of the Grassmann coordinates 
\begin{equation}\label{2.6}
\Theta^\alpha=(1+{i\over{3!R}}\bar\theta\theta)^{1\over 2}\theta^\alpha.
\end{equation}
Then, because of the symmetry 
properties of the Dirac matrices $\gamma^a$
and $\gamma^{ab}$, the Cartan forms become bilinear in $\Theta$
\begin{equation}\label{2.7}
E^a=e^a(x)-i\Theta\gamma^a{\cal D}\Theta,
\end{equation}
\begin{equation}\label{2.8}
\Omega^{ab}=\omega^{ab}(x)+{i\over{2R}}\bar\Theta\gamma^{ab}{\cal D}\Theta.
\end{equation}

\section*{Appendix 2}

\subsection*{The $OSp(1|2n)$ superalgebra and $OSp(1|2n)$ Cartan forms}
The generators of the $OSp(1|2n)$ superalgebra are a 
symmetric bosonic (spin)tensor $M_{\a\b}= M_{\b\a}$  $(\a =1,...,2n)$
and a $2n$--component Grassmann spinor  $Q_\a$, which satisfy the following(anti)commutation relations 

\begin{eqnarray}\label{alg}
 && [M_{\a\b}, M_{\g\d}]=-i\alpha [C_{\g (\a } M_{\b ) \d}
 + C_{\d (\a } M_{\b )\g}],\qquad \nonumber \\ 
 && [ M_{\a\b }, Q_\g  ] =-i\alpha C_{\g ( \a  } Q_{\b )}, \qquad \\
 && \{ Q_{\a}, Q_\b \} = M_{\a\b}, \qquad \nonumber
\end{eqnarray}
where $C_{\a\b}= - C_{\b\a}$  is 
a constant $2n\times 2n$ antisymmetric matrix (symplectic metric). 
Note that to have the correspondence with the form of $OSp(1|4)$
superalgebra \p{2.1} the factor $\alpha$ should be chosen to be
$\alpha={4\over R}$.

When $n=2^{k\over 2}$, $C$ can be regarded as 
a charge conjugation matrix and $Q_\a$ as
a spinor representation of a D--dimensional pseudo-rotation group 
$SO(t,D-t)$ 
with an appropriately chosen number of dimensions $D$  
and time--like dimensions $t$ of space--time.

For instance, when $n=16$ the generators $Q_\a$ of $OSp(1,32)$
can be associated with $SO(1,10)$ Majorana spinors in $D=11$ or
two $SO(1,9)$ Majorana--Weyl spinors of the same or opposite
chiralities in $D=10$. This makes the $OSp(1,32)$ supergroup to be
related to M--theory and superstring theories.
$OSp(1,32)$ is a subgroup of $OSp(1|64)$, and the two supergroups are
extensions of the isometry supergroups $SU(2,2|4)$, $OSp(8|4)$ and 
$OSp(2,6|4)$ of $D=10$ and $D=11$ AdS superspaces  \cite{van,ferr,bars}.

 From a perspective of $D=11$ supergravity and  M--theory the $OSp(1|32)$ 
superalgebra contains 
the $SO(1,10)$ covariant bosonic generators $P_a, M_{ab}=-M_{ba}$ ~and ~
$M_{a_1\ldots a_5}= M_{[a_1\ldots a_5 ]}$. 
A contraction of $OSp(1|32)$  produces the M--algebra \cite{hp,town} 
with $M_{ab}$ and $M_{a_1\ldots a_5}$ becoming tensorial central charges. 

To compute the $OSp(1|2n)$ Cartan forms we choose the following
parametrization of the $OSp(1|2n)$ supergroup element
\begin{equation}\label{A41}
{\cal G}(y,\theta)=B(y)F(\theta)=B(y)e^{i\theta^\alpha Q_\alpha},
\end{equation}
where remember that $y^{\alpha\beta}=y^{\beta\alpha}$ are 
$Sp(2n)$ coordinates.

The $OSp(1|2n)$ Cartan forms are
\begin{equation}\label{ACf}
{1\over i}{\cal G}^{-1} (y , \th ) d {\cal G} (y , \th ) 
\equiv
 {1\over i} \left[F^{-1} (B^{-1} dB) F + F^{-1} dF\right] 
= {E}^\a Q_\a +  {1\over 2} \Om^{\a\b} M_{\a\b}.
\end{equation}

Let us start with computing the $F^{-1}dF$ term of \p{ACf}.
\begin{equation}\label{CfF}
{1\over i}F^{-1}(\th )dF(\th ) =
\S_{n=0}^{\infty } { i^{n} \over (n+1)!} Ad_{\th Q}^n (d\th Q)
\equiv 
{\cal E}^\a Q_{\a} +
{1\over 2} \Om_1^{\a\b} M_{\a\b},
\end{equation}
where
\begin{equation}\label{Ad}
Ad_{B} A \equiv [A,B]
\end{equation}
To calculate the forms $\Om_1^{\a\b}$
and ${\cal E}^\a$ \p{CfF}, note that 
\begin{equation}\label{ad1}
Ad_{\th Q} (d\th Q) \equiv
[d\th Q, \th Q]
= - d\th^{(\a} \th^{\b )} M_{\a\b }
\end{equation}
\begin{equation}\label{ad2} 
Ad_{\th Q}^2 d\th Q \equiv
[[d\th Q, \th Q], \th Q]
= -i\a d\th^{(\b} \th^{\a )} \th_{\b } Q_{\a }
\end{equation}
$$
Ad_{\th Q}^3 (d\th Q) \equiv  
[[[d\th Q, \th Q], \th Q], \th Q] 
$$
$$
= 
-\left({{i\alpha}\over 2} ~\th^\g \th_\g \right) 
[d\th Q, \th Q]=-\left({{i\alpha}\over 2} 
~\th^\g \th_\g \right)Ad_{\th Q}(d\th Q),
$$
$$
Ad_{\th Q}^4 (d\th Q) \equiv 
[[[[d\th Q, \th Q],\th Q],\th Q],\th Q]
$$
$$
 = -\left({{i\alpha}\over 2} ~\th^\g \th_\g \right) 
[[d\th Q, \th Q],\th Q]
$$
$$
=-\left({{i\alpha}\over 2} ~\th^\g \th_\g \right)Ad_{\th Q}^2(d\th Q). 
$$
Thus we arrive at the recursion relation  
\begin{equation}\label{Ad2Q}
Ad_{\th Q}^{l + 2} (d\th Q) 
= -\left( {{i\alpha}\over 2} ~\th^\g \th_\g \right) 
Ad_{\th Q}^{l} (d\th Q) \qquad for \quad l\geq 1
\end{equation}
and can express 
all higher commutators through either \p{ad1} or \p{ad2}   
multiplied by a corresponding power of 
$\left({{i\alpha}\over 2} ~\th^\g \th_\g \right)$. 

In such a way we arrive at the generic expression
for the forms \p{CfF}
\begin{equation}\label{calEa}
{\cal E}^{\a} = d\th^\a +i
d\th^{(\a}\theta^{\b)}\th_\beta  ~
\S_{l =0}^{{n}-1 } {\alpha\over {(2l+3)!}} 
\left( {{i\alpha}\over 2} ~\th^\g \th_\g \right)^{l} 
\end{equation}
\begin{equation}\label{Om1}
\Om_1^{\a\b} = - id\th^{(\a}\th^{\b)} 
\S_{\l =1}^{{n}-1} {1\over (2l+2)!} 
\left({{i\alpha}\over 2} ~\th^\g \th_\g \right)^{l}
\end{equation}

To calculate the first term in \p{ACf} 
\begin{equation}\label{CfFBBF}
{1\over i}F^{-1} (B^{-1} dB) F = 
F^{-1} ({1\over 2} \om^{\a\b} M_{\a\b} ) F \equiv 
E^\a_0 + {1\over 2} \Om_0^{\a\b} M_{\a\b}  
\end{equation}
we note that because $\om^{\a\b}(y)$ is symmetric 
the following relation holds 
\begin{equation}\label{id1}
\th_\b \th^\g \om_\g^{(\a } \th^{\b )}= 
{1\over 2} ~\th^\g \th_\g ~ (\th \om )^{\a }.  
\end{equation}
Then one finds
\begin{equation}\label{Ad2M}
Ad_{\th Q}^{l + 2} {1 \over 2} (\om M)  
= -\left({{i\a}\over 2} ~\th^\g \th_\g \right) 
Ad_{\th Q}^{l} {1 \over 2}(\om M)   \qquad for \quad l\geq 1
\end{equation}
Using \p{Ad2M} we get
the following expressions for the forms \p{CfFBBF}
\begin{equation}\label{Ea0}
{E}_0^{\a} = 
{ \a \over 2}  (\th \om )^\a 
\S_{l =0}^{{n}-1 } {1\over (2l+1)!} 
\left({{ i\alpha}\over 2} ~\th^\g \th_\g \right)^{l} 
\end{equation}
\begin{equation}\label{Om0}
\Om_0^{\a\b} = 
\om^{\a\b}(y) 
- {{i\alpha}\over 2} (\th \om)^{(\a}\th^{\b )} 
\S_{l =0}^{{n}-1} {1\over (2l+2)!} 
\left({{i\alpha}\over 2} ~\th^\g \th_\g \right)^{l}
\end{equation}

Note that in \p{Ea0} and \p{Om0} the polynomials in $\theta^\g\theta_\g$
are the same  as in  \p{calEa} and \p{Om1}. 
Thus, incerting \p{calEa}, \p{Om1},
\p{Ea0} and \p{Om0} into \p{ACf}  we get the following expressions  
for the $OSp(1|2n)$ Cartan forms.
\begin{equation}\label{AEa(D)}
{E}^{\a} = {\cal D}\th^\a +i{\cal D}\th^{(\a } \th^{\b )} \th_\b 
P_1(\theta\theta), 
\end{equation}
\begin{equation}\label{AOm(D)}
\Om^{\a\b} = 
\om^{\a\b}(y) 
+i \th^{(\a }{\cal D} \th^{ \b )}P_2(\theta\theta),
\end{equation}
where 
\begin{equation}\label{poly1}
P_1(\theta\theta)=\S_{l =0}^{n} {{\alpha}\over {(2l+3)!}} 
\left({{i\a}\over 2} ~\th^\g \th_\g \right)^{l}
\end{equation}
\begin{equation}\label{poly2}
P_2(\theta\theta)=\S_{l =0}^{{n}} {1\over (2l+2)!} 
\left({{i\alpha}\over 2} ~\th^\g \th_\g \right)^{l},
\end{equation}
and 
\begin{equation}\label{ADth}
{\cal D}\th^{\a}= d\th^{\a}+ {\alpha\over 2} \om^\a_{~\b}(y)\th^\b. 
\end{equation}
The polynomial $P_2$ \p{poly2} can
be hidden into rescaled $\Theta=\sqrt{P_2}\theta$, so that
$\Omega^{\a\b}$ become bilinear in Grassmann variables
\begin{equation}\label{Asimple}
\Om^{\a\b} = 
\om^{\a\b}(y) 
+ i \Theta^{(\a }{\cal D} \Theta^{ \b )}.
\end{equation}
It is then not hard to varify (using the Maurer--Cartan equations \p{MC1} and
\p{MC2}) that the odd Cartan forms \p{AEa(D)} take the form
\begin{equation}\label{alsimple}
E^\a=P(\Theta^2){\cal D} \Theta^{\a}-\Theta^{\a}{\cal D}P(\Theta^2),
\quad {\rm where}\quad P(\Theta^2)=\sqrt{1+{{i\a}\over 8}\Theta^\b\Theta_\b}.
\end{equation}

Having in hand the $OSp(1|2n)$ Cartan forms it is straightforward to
get the Cartan forms corresponding to any coset superspace $OSp(1|2n)/H$
with $H$ being a bosonic subgroup of $OSp(1|2n)$. To this end in
\p{Asimple} and \p{alsimple} one should
simply put to zero all parameters $y^{\a\b}$ corresponding to the subgroup
$H$. Then $\omega^{\a\b}$ will depend only on the bosonic coordinates of
the supercoset $OSp(1|2n)/H$, and \p{Asimple} will contain the even
supervielbeins and the spin connection of $OSp(1|2n)/H$.

\end{document}